\documentstyle [11pt,epsf] {article}
\begin{document}
\begin{center}
\Large
{\bf Evidence of L\'{e}vy stable process in tokamak \\
edge turbulence}\\[4ex]
\small
R. Jha$^1$, P. K. Kaw$^1$, D. R. Kulkarni$^2$, J. C. Parikh$^2$, and ADITYA Team$^1$\\
\begin{enumerate}
\item Institute for Plasma Research, Bhat, Gandhinagar-382 428, INDIA
\item Physical Research Laboratory, Navarangpura, Ahmedabad- 380 009, INDIA
\end{enumerate}
\normalsize

\vspace{4ex}
\baselineskip = 0.75 cm
{\bf ABSTRACT}\\[2ex]
\end{center}
In an effort to understand the fundamental physics of turbulent transport of particles
and heat in a tokamak, the floating potential fluctuations in the the scrape-off layer
plasma of ohmically heated ADITYA tokamak are analysed for self-similarity using 
distribution function approach. It is observed that the distribution function
of a sum of $n$ data points converges to a L\'{e}vy
distribution of scale index, $\alpha$ = 1.111 for $n \le $ 40
and $\alpha$ = 2.0 for larger $n$. In both scaling ranges, the edge fluctuation
is self-similar. This observation is backed by several supporting evidences. 
The results
indicate that the small scale fluctuations transport matter and heat dominantly
 by convection 
whereas the transport due to large scale fluctuations is by a  diffusive process.\\[8ex]
\newpage
Experimental observations [1] and numerical simulations [2] have shown that random
fluctuations in the edge plasma of a tokamak exhibit a tendency to organize themselves into
coherent structures. These structures are expected to play an important role in causing
transport of particles, a topic of considerable interest in
modern fusion devices. The measurement of fluctuation induced particle flux shows
bursty nature [3]. An attempt has been made to explain the bursty nature of particle
transport in terms of self organized criticality (SOC) as indicated by the observation
of long range time correlation and self similarity in the fluctuation data [4].
However, whether or not the long range correlation is a manifestation of SOC
(and a breakdown of the standard transport paradigm) is an open question [5].\\

In this paper, we carry out an additional investigation of bursty nature of particle
transport in tokamak in terms of L\'{e}vy flights. The paradigm of L\'{e}vy flight has been 
used recently to explain anomalous transport in such diverse subjects as fluid
dynamics [6], transient polymers [7], subrecoil laser cooling [8], human heart beat
[9], movement of stocks in financial markets [10,11] and fluid turbulence [12].
The L\'{e}vy scale index, $\alpha$ is
intimately related to the index, $\nu$ of the transport equation : $<R^2(t)> = D t^{\nu}$,
where $<R^2(t)>$ is the mean square distance travelled by the test particle in the random
field in time $t$ and $D$ is the transport coefficient. For a diffusive transport,
$\nu = 1$ whereas $\nu > 1$ indicates anomalous transport. In general, $\alpha = 2/\nu$ and
hence $0 \le \alpha <2$ represents anomalous transport [13]. To demonstrate L\'{e}vy
process and the resultant self- similarity, we have followed the approach of probability
distribution function (PDF) [10,11]. This approach is superior because the PDF involves 
the entire range of fluctuation amplitude unlike the rescaled range (R/S) analysis
which essentially describes the two point function of the PDF [4]. 
We provide evidence of a self- similar non-Gaussian process,
which is a L\'{e}vy process, over a range of time scales up to 3 times the correlation
time ($\tau_{ac}$). At longer time scales the process is Gaussian. The observation 
implies a new paradigm of turbulent transport in tokamaks, viz., anomalous  transport 
at short time scales and diffusive transport 
at larger time scales.\\

It has been shown by L\'{e}vy [14] that the sum of $n$ independent stochastic variables,
with a probability distribution having power-law wings, converges to a stable process
characterized by the L\'{e}vy distribution. 
The process is stable because the PDF of the sum, 
$Z_n = \sum_{i=1}^n X_i$, of the stochastic  variables
 $\{X\}$, has the same functional form as the PDF of $X_i$.
The PDF of a symmetrical Levy stable process is given by [15,16]
\begin{equation}
p(Z_n) =
\frac{1}{\pi}\int_{0}^{\infty}\exp(-n\gamma
q^{\alpha})\cos(qZ_n)dq
\end{equation}
where $\alpha$ $(0 < \alpha \leq 2)$ and  $\gamma(>0)$ are the
scale index and scale factor respectively. If the central region of the distribution is
well described by a L\'{e}vy stable process, then the  
the probability of return to the origin is
given by [10]:
\begin{equation}
p(Z_n =0) = \frac{\Gamma(1/\alpha)}{\pi \alpha
(n\gamma)^{1/\alpha}}
\end{equation}
where $\Gamma$ is a Gamma function. The L\'{e}vy stable
symmetrical distributions rescale under the following
transformations:
\begin{equation}
Z_s = \frac{Z_n}{n^{1/\alpha}}
\end{equation}
and
\begin{equation}
p_s(Z_s) = n^{1/\alpha} p(Z_n)
\end{equation}
\vspace*{0.3cm}
Although L\'{e}vy stable process is characterized by infinite variance, the existence 
of finite variance in physical systems is treated either by introduction of spatiotemporal 
coupling [13] or by means of truncated L\'{e}vy distribution (TLD) where an exponential tail
is imposed on the values of the stochastic variable [15,16]. The TLD is 
a L\'{e}vy stable process.  
In this paper, we report the evidence of a L\'{e}vy
stable process in the turbulent edge plasma of ohmically heated ADITYA tokamak.\\
 
For this experiment, ADITYA tokamak is operated with with
the following discharge parameters: plasma current, $I_p$ = 64$\pm$4 kA,
toroidal magnetic field at axis, $B_T$ = 0.75 T, chord averaged
plasma density, $\bar{n_e} = 1\times 10^{19} m^{-3}$, major radius
,$R_0$ = 0.75 m, minor radius, $a$= 0.25 m. The floating potential
is measured using a poloidal array of Langmuir probes made up of
molydenum wires (length = 3.5 mm, diameter= 1 mm, separation= 3
mm) and located in the Scrape-off layer plasma 3 mm behind the
limiter. The probes on a movable shaft are mounted on the top
port, toroidally $72^o$ away from the limiter in the electron side. The data are
aquired at a sampling rate of 1 MHz in the time window 25-49 ms.
The mean potential lies between -4 to 1 volts.
The fluctuation data are generated after removing the 1 ms running
average. With this procedure, we expect to remove the very low frequency fluctuations
which are not part of plasma turbulence. A stationary segment of 
data during 35-49 ms are chosen
from seven discharges (8746-48 and 8752-55) for further
analysis. The autocorrelation time, $\tau_{ac}$= 12 $\mu$s. The
autocorrelation function falls sharply and crosses zero at about 
$t/\tau_{ac} = 2$ with no evidence of algebric tail. \\

 For further analysis, stationary
fluctuation time series from 7 discharges are normalized with
their root mean square (RMS) amplitude and stacked together. The
distribution function is obtained using 980000 data points. This
is shown in Fig.(1), together with a Gaussian having the same
variables. The PDF of the fluctuation is nearly symmetrical
(skewness = -0.38) but has a non-neglible kurtosis (1.24). The evidence
of high kurtosis is seen in the distribution peaked near the origin. We next
try to fit the PDF of the fluctuations data to a L\'{e}vy
distribution. \\

In order to determine the L\'{e}vy scale index
$\alpha$, the number of convolutions is chosen in the logarithmic
intervals, $n= 2^{i-1}$, where $i=$1,2, ... 11. The original time series is
divided into non-overlapping blocks of $n$ data points and a new variable,
$Z_n$ is generated for each block. The new time series is then subjected to
L\'{e}vy analysis. The probability of
return to origin, $p(Z_n = 0)$ is plotted as a function of 
$n$. This is done for the following reasons: (i) the
distribution function is  most accurate at $Z_n=0$, and (ii) the analytical form
of only $p(Z_n=0)$ is known [Eq.(2)]. 
It is observed that on a log-log plot (Fig. 2),
there are two clear slopes with values -0.9 and -0.5 in the
$n$-ranges 1-40 and 40-500 respectively. The slope with value
-0.9 corresponds to L\'{e}vy index $\alpha$ = 1.111 and that with
value -0.5 to $\alpha$ = 2.0.
The values of the scale
factor, $\gamma$ in the two scaling ranges are 0.64$\pm$0.06 and 8.03$\pm$0.76 
respectively. It is important to note that the cross-over to a Gaussian process
takes place at $n=n_x=$ 40 which is much larger than that expected form the
central limit theorem [17]. \\

In Fig.(3a) we compare the PDF observed for $n$=1
with the L\'{e}vy stable distribution of the scale index,
$\alpha$=1.111 and scale factor, $\gamma$=0.59. The L\'{e}vy
distribution is a good fit in the central region of the PDF upto
3$\sigma$ values of the stochastic variable. Beyond 3$\sigma$, the
PDF  departs from both L\'evy and Gaussian forms. Note that the
values of $\alpha$ and $\gamma$ are obtained in the
self-similarity range of $n$=1-32 $\mu$s. This range represents
small scale fluctuations in the floating potential.
Figure 3(b) shows the full distribution function for the
$n-$values 1, 2, 4, 8, 16 and 32. As the number of convolutions
increases, the distribution function becomes broad and the peak
value (or, the probability of return to origin) decreases. If the
stochastic variable, $Z_n$ and the distribution function, $p(Z_n)$
are rescaled in accordance with Eq.(3) and Eq.(4) respectively,
the distribution functions, $p_s(Z_s)$ collapse on $n=$1
distribution (Fig. 3(c)). Thus, the distribution functions of
different convolutions are self-similar in the range $n=$1 to 32.
A self-similarity is also obtained for $n-$values 64, 128, 256 and
512 by using $\alpha$=2.0 and $\gamma$=8.03$\pm$0.76. Figure 4 shows
the rescaled distribution functions in the two scaling regimes
on the same graph.
The difference of $\log(p_s(Z_s=0))$ values in the first and the second 
scaling regimes provides a measure of the 'distance', 
$\Delta$ between the two scaling regimes [15].
The $\Delta \approx$ 0.6 for the results shown in Fig. 4 indicates that the 'distance'
is significantly large. \\

We have verified the above results by carrying out the following analysis:
(i) the root mean square (RMS) amplitude of the stochastic variable ($Z_n$),
$\sigma_{Z_n}$ shows a power-law behaviour, $\sigma_{Z_n} \sim n^{\nu/2}$.
The exponent, $\nu$=1.8 in the short time scaling range ($n$=1-32) whereas,
$\nu$=1.0 in the long time scaling range ($n$=100-800). Thus, the small scale
fluctuations show non-Gaussian PDF and a self-similar scaling behaviour.
The long scales show self-similarity of Gaussian type. It can be argued that
the RMS amplitude of potential fluctuation is proportional to the root mean square distance
travelled by a test particle in the random field [18].
(ii) the rescaled range ($R/S$) analysis on the combined data set shows a
somewhat longer scaling range at short time scales ($n$= 1-128) having Hurst
parameter, $H$=0.9. At long time scales, $n$=256-4096, the value of $H$ = 0.6.
One is tempted to interpret it as an evidence of long range self-smilarity
of non-Gaussian type. However, it should be pointed out that $R/S$ range analysis
has tendency to give somewhat higher values of $H$ [19]. 
Therefore, we consider that
at long time range, the fluctuation data show self-similarity of Gaussian type.
(iii) when we calculate the kurtosis ($K$) of the time series corresponding to different
convolutions ($n$), it is observed that kurtosis decreases with increasing $n$ and
$K <$ 0.2 for $n >$ 60. This result is similar to those reported earlier in
tokamak [20] and stellerator [21]. Thus, the small scale fluctuations are non-Gaussian 
whereas the
large scales are Gaussian.\\

In conclusion, we have presented evidences of a L\'{e}vy stable process in tokamak
edge turbulence. The sum of $n$ data points of a fluctuation time series
converges to a L\'{e}vy distribution of scale index, $\alpha$=1.111 for $n \le 40$
and $\alpha$= 2.0 for larger $n$. The probability distribution functions are 
self similar in both scaling ranges. This observation is backed by several
supporting evidences.\\

We finally give a possible interpretation of these observations and speculate on their
implication for the problem of bursty transport due to turbulence in the tokamak
edge region. We have earlier carried out conditional statistical analysis of Langmuir probe
data in the edge of tokamak ADITYA (for similar discharges) and shown that the fluctuations
are dominated by coherent structures which appear intermittently in space and time,
typically lasting 25-30 $\mu$s and having poloidal scale lengths of a few cm. These
structures are also associated with sharp radial potential 
gradients separating the last closed 
magnetic surface from the scrape-off layer plasma. Bursts of turbulent transport into the
scrape-off layers are related to intermittent breakdown of radial confinement and the
coherent structures extending into the open field line regions. We believe that the 
non-gaussian PDFs of potential fluctuation ($\phi$) described above are due to such coherent structures. For time
scales longer than the life time ($\approx$ 25-30 $\mu$s), we observe only a randomized average behaviour and hence infer only Gaussian statistics and diffusive behaviour. For shorter
time scales, on the other hand, convective effects due to coherent structures dominate and
we observe the anomalous behaviour, $ <R^2> \sim t^{\nu}$ with $\nu$=1.8. We may also make
an estimate of the enhancement of turbulent transport due to convective effects
introduced by the presence of coherent
eddy-like structures by using the analysis of Rosenbluth et {\it al.} [22]. The enhancement
factor $f = D^*/D \sim P^{1/2}$ where the Peclet number $P \sim vd/D$ is the ratio of the
diffusion time to the eddy turn-over time. For a typical eddy size of $d \sim 1$ cm, the 
typical eddy velocity, $v \sim cE/B \sim 10^5$ cm/s and diffusion coefficient, 
$D \sim 10^4$ cm$^2$/s [1,23], we find an enhancement factor $f \sim 3$. 
Thus, the presence of
transient  coherent eddies can give sudden avalanche- like enhancements of 
transport by a factor of 
3 or more, resulting in bursty nature of transport. 

\newpage 
\begin{center}
{\bf REFERENCES}\\
\end{center}
\begin{itemize}
\item[(1)] B. K. Joseph et al., Phys. Plasmas {\bf 4}, 4292 (1997); S. Benkadda et al.,
	Phys. Rev. Lett. {\bf 73}, 3403 (1994).
\item[(2)] A. E. Koniges et al., Phys. Fluids B {\bf 4}, 2785 (1992); 
	   J. A. Crotinger and T. H. Dupree, Phys. Fluids B {\bf 4}, 2854 (1992).
\item[(3)] M. Endler et al. Nucl. Fusion {\bf 35}, 1307 (1995).
\item[(4)]  B. A. Carreras et al., Phys. Rev. Lett. {\bf 80}, 4438 (1998);
	  B. A. Carreras et al., Phys. Plasmas {\bf 5}, 3632 (1998).
\item[(5)]  J. A. Krommes and M. Ottaviani, Phys. Plasmas {\bf 6}, 3731 (1999).
\item[(6)] T. H. Solomon et al., Phys. Rev. Lett. {\bf 71}, 3975 (1993).
\item[(7)] A. Ott et al., Phys. Rev. Lett. {\bf 65}, 2201 (1990).
\item[(8)] F. Bardou et al., Phys. Rev. Lett. {\bf 72}, 203 (1994).
\item[(9)] C. K. Peng et al., Phys. Rev. Lett. {\bf 70}, 1343 (1993).
\item[(10)]  R. N. Mantegna and H. E. Stanley, An Introduction to Econophysics,
        Cambridge University Press, U.K. (2000), p. 26.
\item[(11)] R. N. Mantegna and H. E. Stanley, Nature {\bf 376}, 46 (1995); 
	S. Ghashghaie et al., Nature {\bf 381}, 767 (1996);
\item[(12)] M. F. Shlesinger et al., Phys. Rev. Lett. {\bf 58}, 1100 (1987).
\item[(13)]  A. Blumen, G. Zumofen and J. Klafter, Phys. Rev. A {\bf 40}, 3964 (1989).
\item[(14)] P. L\'{e}vy, Th\'{e}orie de l\'Addition des Variables Al\'{e}atoires
		(Gauthier-Villars, Paris, 1937).
\item[(15)]  R. N. Mantegna and H. E. Stanley, Phys. Rev. Lett. {\bf 73}, 2946 (1994);
		R. N. Mantegna, Phys. Rev. E {\bf 49}, 4677 (1994).
\item[(16)] J.-P. Bouchaud and M. Potters, Theory of financial risk: from data
		analysis to risk management, Cambridge University Press, U. K. (2000).
\item[(17)] For a time series, $X_i = cos(\theta_i)$, where $\theta_i$ is a uniformly
distributed random phase in the range $0- 2\pi$, the sum, $Z_n = \sum_{1}^{n}X_i$ becomes
Gaussian for $n \ge 8$. 
\item[(18)] The tokamak edge plasma is dominated by low frequency electrostatic turbulence
in which the radial displacement ($R$) is given by ${\bf E} \times {\bf B}$ motion, viz.,
$dR/dt=(1/RB)\partial \phi/\partial \theta$, where $B$ is the magnetic field, $\phi$ is the 
electric potential. Noting further that the plasma has mean
poloidal velocity of order $v_0$, we find $R \approx \phi/v_0B$ indicating the potential 
$\phi$ is a direct measure of radial displacement. 
\item[(19)] B. A. Carreras et al., Phys. Plasmas {\bf 6}, 1885 (1999).
\item[(20)] R. Jha et al., Phys. Rev. Lett. {\bf 69}, 1375 (1992);
		 R. Jha and Y. C. Saxena, Phys. Plasmas {\bf 3}, 2979 (1996).
\item[(21)]   V. Carbone et al., Phys. Plasmas {\bf 7}, 445 (2000).
\item[(22)] M. N. Rosenbluth et al., Phys. Fluids {\bf 30}, 2636 (1987). 
\item[(23)] R. Jha et al., Nucl. Fusion {\bf 33}, 1201 (1993).
\end{itemize}

\newpage
\begin{center}
{\bf FIGURE CAPTION}
\end{center}
\begin{enumerate}
\item[(1)] The probability distribution function (PDF) of floating potential fluctuation.
The amplitude ($Z$) is normalized to the standard deviation. The symbol (open circle)
shows the experimental data points whereas the dotted line shows a Gaussian PDF of unity
standard deviation.\\ 
\item[(2)] The probability of return to the origin, $p(Z_n = 0)$ as a function of 
 the number of data points ($n$) in the non-overlapping blocks. For $n$=1024 and beyond,
the PDF is not well defined and $p(Z_n=0)$ estimates may be in error. Since we have filtered
out time scales larger that 1 ms, they do not contribute to the PDF.\\
\item[(3)] (a) The PDF, $p(Z)$ for the experimeental data (bullets) together with the Gaussian distribution 
(dotted line) and the L\'{e}vy distribution (solid line), (b) the PDF, $p(Z_n)$ vs.
$Z_n$ for different $n$. The width of $p(Z_n)$ increases with
increasing $n$, (c) the rescaled PDF, $p_s(Z_s)$ for the first six convolutions ($n=$1-32).\\ 
\item[(4)] The comparision of rescaled PDF, $p_s(Z_s)$ in the scaling ranges,
$n=$1-32 (solid lines, without symbol) and $n=$64-512 (with symbols).
\end{enumerate}
\newpage
\begin{center}
{\bf Figure 1}
\end{center}
\begin{center}
\leavevmode
\epsfxsize 4in
\epsfysize 5in 
\epsfbox{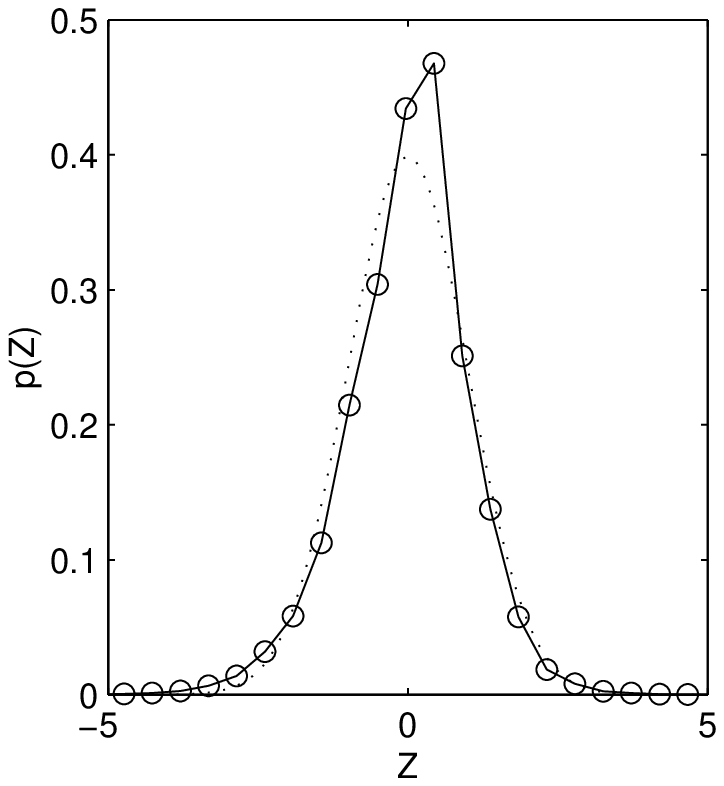}
\end{center}
\newpage
\begin{center}
{\bf Figure 2}
\end{center}
\begin{center}
\leavevmode
\epsfxsize 4in
\epsfysize 5in
\epsfbox{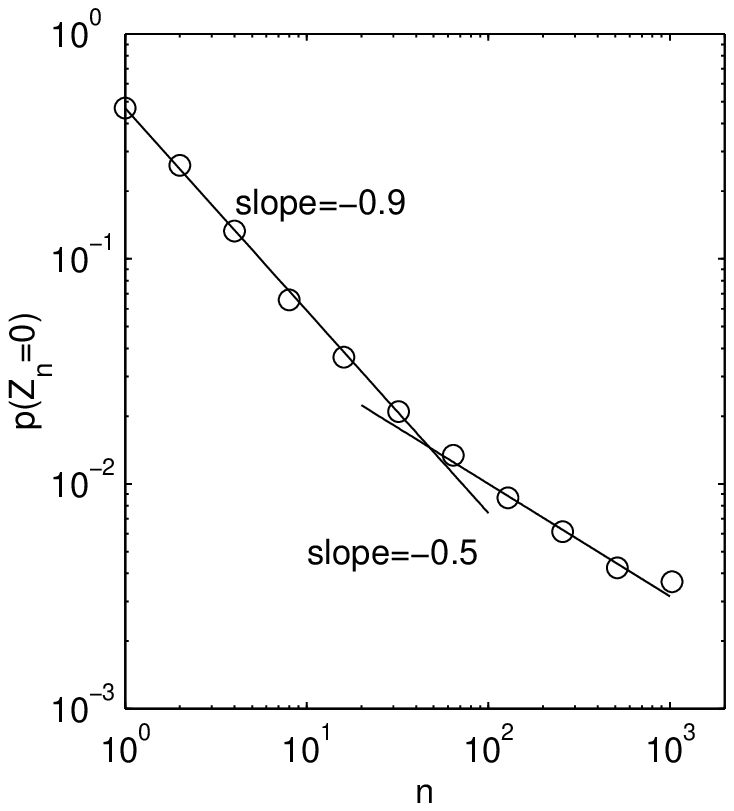}
\end{center}
\newpage
\begin{center}
{\bf Figure 3}
\end{center}
\begin{center}
\leavevmode
\epsfxsize 4in
\epsfysize 5in
\epsfbox{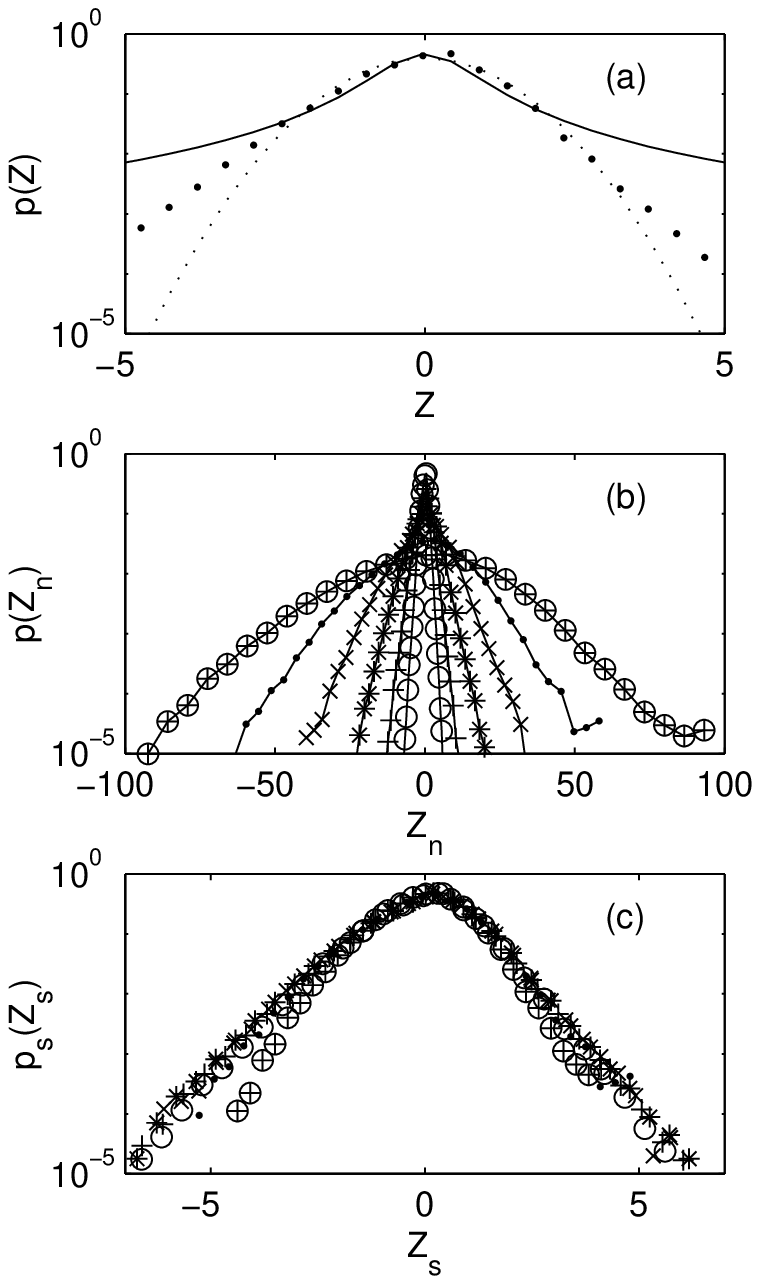}
\end{center}
\newpage
\begin{center}
{\bf Figure 4}
\end{center}
\begin{center}
\leavevmode
\epsfxsize 4in
\epsfysize 5in
\epsfbox{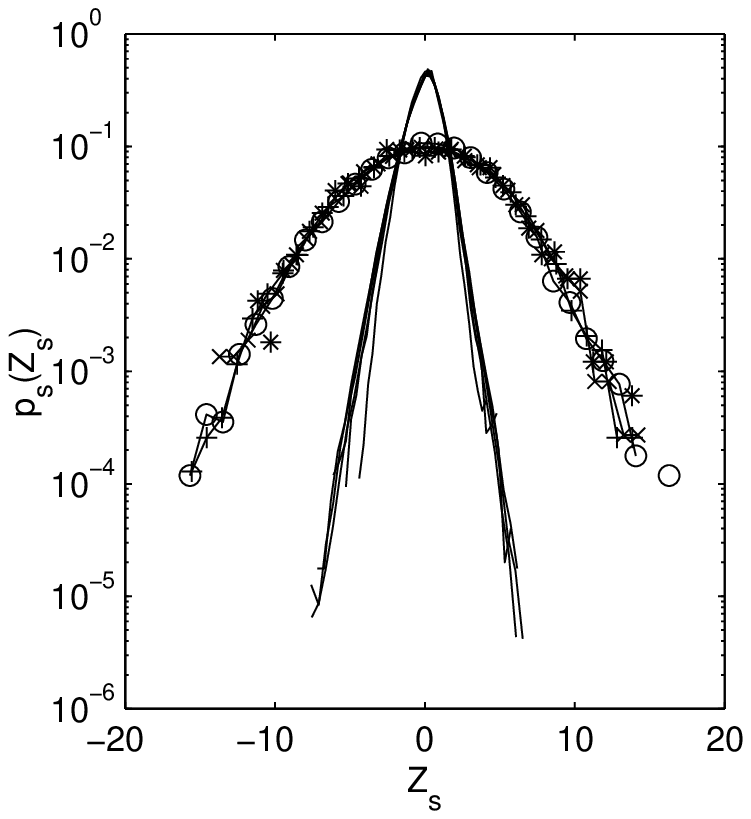}
\end{center}
\end{document}